\documentstyle[12pt,cite,rotate,epsf]{article}

\newcommand{\equ}[1]{Eq.\,(\ref{#1})}

\newcommand{\ew}{electroweak~}
\newcommand{\non}{\nonumber}
\newcommand{\gsim}{\;\rlap{\lower 3.5 pt \hbox{$\mathchar \sim$}} \raise 1pt
 \hbox {$>$}\;}
\newcommand{\lsim}{\;\rlap{\lower 3.5 pt \hbox{$\mathchar \sim$}} \raise 1pt
 \hbox {$<$}\;}

\newcommand{\msbar}{\overline{\rm MS}}

\newcommand{\smallz}{{\scriptscriptstyle Z}} 
\newcommand{\smallw}{{\scriptscriptstyle W}} %
\newcommand{\smallh}{{\scriptscriptstyle H}} %

\newcommand{\mz}{M_\smallz}
\newcommand{\mw}{M_\smallw}

\newcommand{\mh}{M_\smallh}
\newcommand{\mt}{M_t}
\newcommand{\mtbar}{\overline{M}_t}

\newcommand{\bb}{B^0-\bar {B}^0}

\newcommand{\eps}{\epsilon}

\def\pl#1#2#3{{\it Phys. Lett. }{\bf B#1~}(19#2)~#3}

\def\pr#1#2#3{{\em Phys. Rev. }{\bf D#1~}(19#2)~#3}
\def\np#1#2#3{{\em Nucl. Phys. }{\bf B#1~}(19#2)~#3}

\newcommand{\be}{\begin{equation}}
\newcommand{\ee}{\end{equation}}
\newcommand{\een}{\end{subequations}}
\newcommand{\ben}{\begin{subequations}}
\newcommand{\beq}{\begin{eqalignno}}
\newcommand{\eeq}{\end{eqalignno}}
\newcommand{\bea}{\begin{eqnarray}}
\newcommand{\eea}{\end{eqnarray}}

\newcommand{\fsl}{\hspace{-0.5em}/}      

\begin{document}
\boldmath
\unboldmath


\thispagestyle{empty}
\rightline{TUM-HEP-331/98}
\rightline{October 1998}
\vspace*{1.2truecm}
\bigskip

\centerline{\LARGE\bf  Electroweak  effects  in  the $B^0 -
\bar{B}^0$ mixing\footnote{Supported by the Bundesministerium f{\"u}r Bildung und
Forschung under contract 06 TM 874 and by the DFG project Li 519/2-2.}}
\vskip1truecm
\centerline{\large\bf Paolo Gambino, Axel Kwiatkowski, and Nicolas Pott}
\bigskip
\centerline{\sl  Technische Universit{\"a}t M{\"u}nchen, Physik-Department}
\centerline{\sl D-85748 Garching, Germany}
\vspace{1.5truecm}
\centerline{\bf Abstract}
 We compute analytically the complete \ew two-loop corrections to the
$\bb$ mixing. These corrections fix the
normalization of the electroweak coupling employed in the
extraction of $|V_{td}|$ and reduce the theoretical uncertainty due
 to higher order \ew effects
from  several percent  to a few parts in a thousand.
If the LO result is expressed in terms of
 $G_\mu$ or of the $\msbar$ coupling $\hat{g}(\mz)$,
the two-loop corrections are $O(1\%)$, the exact value depending on the mass 
of the Higgs boson.
We discuss in detail the renormalization procedure and the scheme and scale
dependence, and provide  practical formulas for the numerical implementation 
of our results.
We also consider the  heavy top mass expansion 
and show that in the case at hand it converges very slowly.
\vspace*{2.0cm}


\newpage


\section{Introduction}
The $\bb$ system offers rich possibilities for studies of CP violation
and the quark mixing structure of the Standard Model 
(see \cite{buras:97} for a
recent review). The physics of this
system is well described by ${\cal H}_{eff}^{\Delta B=2}$, 
the effective low-energy Hamiltonian for the $B^0
\leftrightarrow \bar B^0$ transition, and the most important
observable directly linked to ${\cal H}_{eff}^{\Delta B=2}$ is
 $\Delta M_{B^0}$, the mass
difference between the heavy and the light mass eigenstates in the $\bb
$ system. Theoretically, this quantity is given by 
\be \label{deltamb}
\Delta M_{B^0}={1 \over m_B} \vert
\langle \bar B^0 \vert {\cal H}_{eff}^{\Delta B=2} \vert B^0 \rangle \vert,
\ee
and the experimentally measured value is \cite{PDG98} 
\be
\Delta M_{B^0} =(0.46 \pm 0.02 )
\times 10^{12}\, s^{-1}.
\ee
If all other ingredients in the evaluation of the r.h.s. of \equ{deltamb}
 are sufficiently well known, one can extract from this
 measurement the absolute
value of the Cabibbo-Kobayashi-Maskawa (CKM) parameter $V_{td}$, which
plays
 an important role in  the
standard analysis of the unitary triangle \cite{buras:97}. This
example illustrates the phenomenological relevance of a precise
determination of ${\cal H}_{eff}^{\Delta B=2}$ and its matrix element, especially
 if one takes into
account the experimental accuracy already achieved in the measurement
of $\Delta M_{B^0}$.

The two main steps for such a determination are (i) the
calculation of ${\cal H}_{eff}^{\Delta B=2}$
 (or more precisely: the corresponding
Wilson coefficient) in perturbation theory -- in particular, this
calculation yields the complete dependence of ${\cal H}_{eff}^{\Delta
B=2}$  on all
the heavy degrees of freedom, and (ii) the evaluation of the
remaining low-energy matrix element on the lattice or by some other
non-perturbative method. Presently, this second step causes 
the largest theoretical uncertainty in the determination
of $\langle {\cal H}_{eff}^{\Delta B=2} \rangle$,
 about $\pm 20 \%$ (dominated by
systematics of the lattice calculation, i.\,e.\ extrapolation to
the continuum  \cite{bernard:98,flynn:97}).
 There is, however, well-founded hope that with
increasing computer power or  by new developments in lattice
theory, this uncertainty may be significantly reduced 
in the near future.

The perturbative calculation, on the other hand, has been by now  performed
at the level of next-to-leading order QCD \cite{buras:90}. This
analysis includes the matching at $O(\alpha_s)$ and the $O(\alpha_s^2)$
anomalous dimension of the relevant four-quark operator; the
achieved accuracy 
is better than $\pm 1 \%$ \cite{buchalla:96}.

\begin{figure}
\centerline{
\epsffile{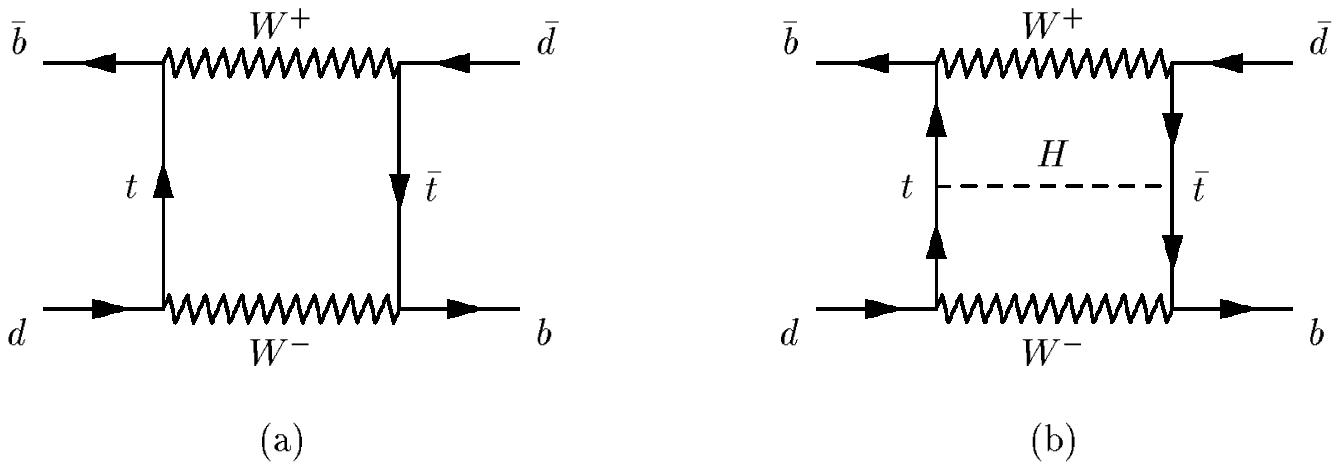}}
\caption{\sf (a) shows one of the box diagrams contributing to the $B^0
\rightarrow \bar B^0$ transition at leading order, whereas (b)
depicts an example of an electroweak two-loop correction to this
process which may be seizable in the large $M_t$ limit.}
\end{figure}

In this paper, we focus on a different part of the perturbative analysis
which has not yet  been considered in the literature: the two-loop
{\em electroweak} correction to the effective Hamiltonian ${\cal
H}_{eff}^{\Delta B=2}$. In our opinion, at least three
 reasons suggest such a  calculation:
\begin{itemize}

\item {\em Reduction of scheme dependence}. The leading order (LO) result
(Fig.\,1a shows one of the relevant diagrams) is proportional to
$g^4$, where $g$ is the weak coupling. However, at LO no renormalization
prescription for $g$ is needed, so one can use the numerical value of
$g$ in {\em any}  renormalization scheme.
For example,  the difference
between $g^4$ at the scale $\mz$ calculated from $\sin^2\theta_{lept}^{eff}
=0.23155$ \cite{LEP} and $\alpha(\mz)=(128.9)^{-1}$ \cite{Jeger}, or from 
the relation $g^2=8 \mw^2 G_\mu/\sqrt{2}$
amounts to about $2.5 \%$. As noted in \cite{buchalla:98},
such ambiguity reflects the
uncertainty of the LO result due to the uncalculated electroweak
two-loop correction, and it already exceeds the existing perturbative QCD
uncertainty mentioned above.
In a similar way, if we express the LO result in terms of
the $\msbar$ coupling $\hat{g}(\mu)$, and change the scale $\mu$ between 
$\mw/2$ and $2\mw$, we obtain a 5\% variation, a normalization ambiguity which
is almost completely removed by the consideration of 
 two-loop \ew effects.

\item {\em Possible large corrections due to a heavy top.} The
coupling of the Higgs particle to the top quark is proportional to
$g\, \mt/\mw$. Since $g\, \mt/\mw \approx  g_s$ at
the scale $\mz$, from diagrams like the one in Fig.\,1b
one can expect large contributions  of roughly the same order of magnitude
as the QCD corrections.

\item {\em Performing a complete two-loop \ew calculation.}
Despite recent efforts, there exist very few nearly complete 
electroweak calculations at the two-loop level \cite{anom_muon}, and many
available results  rely on heavy mass expansions \cite{barb,DGV,DGS,cks},
 which are known to work very well in specific examples
\cite{cks,weiglein}. 
In other cases (e.\,g.\ the important two-loop QED corrections to muon decay
\cite{stuart} and some   \ew   corrections to $B \rightarrow
X_s \gamma$ computed in \cite{czarnecki:98}),  only
 gauge-invariant subsets of diagrams have been  considered.
 Similarly to the case of the anomalous magnetic 
moment of the muon considered in \cite{anom_muon}, the mixing 
$\bb $ is a low-energy process which first occurs at one-loop level. 
Its two-loop \ew corrections 
therefore require only  one-loop renormalization and the relevant two-loop 
integrals can be expressed in terms of elementary  functions, because
 the external momenta can be generally neglected.
As will be illustrated in the following, this allows one to compute 
analytically the complete 
electroweak effects in a relatively simple way.
The calculation presented here has therefore some
interest in itself, independent of phenomenological applications. In
particular, it exhibits the complete Higgs mass dependence and
enables one to assess the validity of the heavy mass expansion
in the case of \ew boxes. Interestingly, our
results show a very slow convergence of the heavy top 
mass expansion in the case at hand.
\end{itemize}

In addition to the strong motivation provided by the above three points, 
it should also  be mentioned that our results represent an important subset of
the \ew corrections to the $K^0-\bar K^0$ mixing, whose treatment is 
more involved and will not be considered in the present
work.

This paper is organized as follows: in Sec.\,2 we outline the strategy we have
devised and the general features 
of the calculation; in Sec.\,3 we describe the renormalization procedure 
in detail, giving explicit expressions for the renormalization constants
we have used.  Finally, Sec.\,4  is devoted to a 
discussion of our results 
and of their main consequences, which are summarized in the  Conclusions.


\section{Outline of the calculation}

\subsection{The effective Hamiltonian}
The effective Hamiltonian ${\cal H}_{eff}^{\Delta B=2}$ for the $B^0
\leftrightarrow \bar B^0$ transition can schematically be written as
\cite{buchalla:96}
\be \label{effhamiltonian}
{\cal H}_{eff}^{\Delta B=2}={1 \over 16 \pi^2} {g^4 \lambda_t^2 \over
32 \mw^2} \, C_{B\bar B}(\{M\},\mu) \, \hat Q_{LL} \,+ \,H.c.,
\ee
where $\lambda_t =V_{tb}^* V_{td}$ denotes the CKM factor, 
$\hat Q_{LL}=\bar b \gamma^\mu (1-\gamma_5) d \otimes \bar b \gamma_\mu
(1-\gamma_5)d$  is the relevant four-quark
operator, and $C_{B \bar B}(\{M\},\mu)$ is the corresponding Wilson
coefficient, to be calculated within perturbation theory. The latter contains
 all information on
the heavy degrees of freedom, generically indicated by $\{M\}$, and depends
additionally on the  scale $\mu$ at which the operator 
$\hat{Q}_{LL}$ is renormalized. In principle, this scale can
be chosen arbitrarily -- physical quantities do not depend on it and
therefore the $\mu$-dependence of the Wilson coefficient
will cancel against an analogous $\mu$-dependence of the matrix
element $\langle \hat Q_{LL} \rangle$ order by order in perturbation theory.
 It is physically natural, however, to choose $\mu$
of the order of  $ m_b$, since the $b$ mass is
clearly the relevant scale for the evaluation of the matrix element
$\langle \hat Q_{LL} \rangle$. 

Including next-to-leading order (NLO) QCD corrections 
as well as two-loop electroweak
corrections, the Wilson coefficient $C_{B \bar B}$ assumes the form
\be
C_{B \bar B}(\{M \},\mu_b) = \bar\eta_{2 B} \,\left\{S_0(w_t)  + \delta S_{ew}
\right\},
\label{effth}
\ee
where the Inami-Lim function ($w_t=\mw^2/M_t^2$)
\be
S_0(w_t)={1-11 w_t+4 w_t^2 \over 4 w_t (w_t-1)^2 }+
{3 \ln w_t \over 2 (w_t-1)^3}, \label{S0function}
\ee
represents the leading order result \cite{inami:81}. The factor
$\bar\eta_{2B}$  contains the complete NLO QCD corrections
including the running of the Wilson coefficient down to the scale $\mu=m_b$
 \cite{buchalla:96}; $\bar\eta_{2B}$ is scheme dependent and its scheme
dependence is  canceled by analogous terms in  the matrix
element (see discussion in \cite{buchalla:96}). In the NDR scheme and using
an $\msbar$ top mass evaluated at $\mt$ (see Sec.\,3.1),   
$\bar\eta_{2B}\approx 0.85$.
By $\delta S_{ew}$, instead, we  denote the electroweak two-loop
correction of order $g^2$. The purpose of this paper is  to calculate
this quantity.
 For a better
understanding of this calculation, and  in order to introduce some
relevant notation, we will first briefly recall some important
features of the leading order (one-loop) calculation.

\subsection{Summary of one-loop results}
\label{sec:oneloop}
An important attribute of Wilson coefficients in general
is their independence of the infrared region of the theory, i.\,e.\ ``soft''
physics. In particular, this means that one can calculate them using
arbitrary kinematic configurations of the external particles (on-shell
as well as off-shell), and arbitrary values for the light quark
masses. In the case at hand, all quarks but the top can be considered
 light, and  only the top has been integrated out in order to obtain
the effective theory of Eq.~(\ref{effhamiltonian}). This freedom
in the choice of the ``soft'' parameters crucially simplifies the
calculation: in the following we will always adopt the simplest
approach and work with zero
external momenta and zero masses of the light quarks (except where
these masses are needed as infrared regulator).

Keeping this in mind, the one-loop amplitude for the process
$\bar b+d \rightarrow b+\bar d$
with vanishing external momenta and all light quark masses set to zero
can be written as
\be
{\cal M}_{1loop}={-i \over 16 \pi^2} {g^4 \over 8 \mw^2} \,\sum_{i,j} 
\lambda_i \lambda_j 
\,{\cal S}^{(i,j)} \,  Q_{LL}
\ee
where $\lambda_i=V_{ib}^* V_{id}$ and 
the sum is over the leading order box diagrams 
with internal quarks $i$ and $j$ ($i,j=u,c,t$), calculated in $n$
dimensions.  ${\cal
S}^{(i,j)}$ is the general box function given e.\,g.\ in Eq.\ (2.2) of
\cite{buras:90}. Using
$m_u=m_c=0$ and the unitarity of the CKM matrix, one obtains
\be \label{oneloopres}
{\cal M}_{1loop}={-i \over 16 \pi^2} {g^4 \lambda_t^2 \over 8 \mw^2}
\left [ {\cal S}^{(t,t)}-2 \,{\cal S}^{(t,c)}+ {\cal S}^{(c,c)} \right ]
\, Q_{LL}.
\ee
The survival of the sole $\lambda_t$ term in \equ{oneloopres}
is a consequence of the {\it hard} (power-like)
GIM cancellation mechanism which takes place in the case of 
\ew boxes. 
For later convenience, we present here the explicit result for the
function
\be \label{Sfunction}
S(w_t,\bar \mu^2/M_t^2) \equiv {\cal S}^{(t,t)}-2 \,{\cal S}^{(t,c)}+ {\cal
S}^{(c,c)} =S_0(w_t)+\epsilon \,
S_1(w_t,\bar \mu^2/M_t^2)+O(\epsilon^2)
\ee
where $\eps= 2-n/2$ and we have included 
the $O(\epsilon)$ terms: $S_0(w_t)$ has already been given in
\equ{S0function}, and the function $S_1$ reads
\bea
S_1(w_t,\bar \mu^2/M_t^2)&=&{3-33 w_t-4 w_t^2 \over 8 w_t (w_t-1)^2}+{2+15 w_t
\over 4 (w_t-1)^3} \ln w_t - {3 \ln^2 w_t \over 4 (w_t-1)^3} \nonumber
\\
&& +
S_0(w_t) \ln {\bar \mu^2 \over M_t^2}, \label{S1function}
\eea
with $\bar \mu^2=4 \pi e^{-\gamma_E} \mu^2$.

Before proceeding further, some notation should be introduced.
 The electroweak coupling is
generically denoted by $g$, the sine and cosine of the Weinberg angle
by $s$ and $c$, respectively. At the order we are working at, however, it
is at some point necessary to specify the renormalization schemes
(to be discussed in the following) in
which these quantities are defined. 
Specifically, a hat ($\hat g$,
$\hat s$, $\hat c$) will always
denote  $\msbar$ scheme quantities  
and the subscript $W$ ($g_\smallw$, $s_\smallw$, $c_\smallw$)
the electroweak on-shell scheme. Furthermore, throughout the paper
the following short-hand forms for the ratios of masses are used:
\bea 
w_t= \frac{\mw^2}{\mt^2}\ ,\ \ \ \ 
z_t= \frac{\mz^2}{\mt^2}\ ,\ \ \ \ 
h_t= \frac{\mh^2}{\mt^2}\ ,
\eea
where $\mw$, $\mz$, $M_t$ and $\mh$ are understood as on-shell masses.

\subsection{The matching procedure at two loops}
We are now ready to describe the strategy we have 
followed to perform our calculation.
As a first step, we compute the complete \ew corrections
to the amplitude, setting all external masses and momenta and the internal
light quark masses to zero.
After renormalization, these corrections are of course ultraviolet finite,
but some of the diagrams containing the photon 
need the introduction of an  infrared regulator.
Following the QCD analysis of \cite{buras:90},
 we choose it to be a common  internal light quark mass.
The renormalized two-loop amplitude can then be expressed as
\be
{\cal M}_{2loop}={-i \over {(16 \pi^2)}^2} {g^6 \lambda_t^2 \over
8 \mw^2} \, \Delta^{(2)} \, Q_{LL}, 
\ee
where $\Delta^{(2)}$ is given  by the sum of (i) the unrenormalized one-particle irreducible two-loop diagrams shown in Fig.\,2a-h, (ii) the various
counterterm contributions $\Delta^c_i$ 
(to be discussed in Sec.~3), (iii) a  contribution $\Delta_{DT}$ 
from the one-particle reducible diagrams depicted in Fig.\,2i:
\be
\Delta^{(2)} = 
\Delta^{unren}_{2loop}+\Delta^c_{M_t}+\Delta^c_{\mw}+\Delta^c_g
+\Delta^c_\psi+\Delta^c_{Tad}+\Delta_{DT} . 
\label{delta}\ee
The last contribution, $\Delta_{DT}$, can be gleaned from \cite{inami:81}; it
originates 
from  a finite but gauge-dependent subset of two-loop diagrams and was 
already considered in \cite{buras:90}. We will discuss it in detail in
Sec.\,2.5.

As a second step, we consider the  effective theory of \equ{effhamiltonian},
in which all  
degrees of freedom of the Standard Model (SM) with $M\geq\mw$ are decoupled, and perform the
matching between the full theory and this effective theory at first order 
in QED. The amplitude in the effective theory is  
\be
{\cal M}_{eff}={-i \over {(16 \pi^2)}^2} {g^6 \lambda_t^2 \over
8 \mw^2} \,\Delta_{eff}^{QED}(\mu) \,Q_{LL}\, ,
\label{deltaeff}
\ee
where\footnote{In calculating 
\equ{qedeff} we have employed the projection
method described in the next subsection, 
 thus conforming to the same choice of evanescent operators
as in \cite{buras:90}. A detailed discussion of the definition of 
evanescent operators can be found in \cite{dugan,herrlich:95}. } 
\be
\Delta^{QED}_{eff}(\mu)= {2 \over 3} s^2 
 \,S_0(w_t) \, \left( \ln q_t- \ln {\mu^2 \over M_t^2}-2 \right ).
\label{qedeff}
\ee
Here, $q_t=m_q^2/M_t^2$ ($m_q$ denotes the generic mass of the internal light
quarks) plays the role of an infrared regulator. 
The electroweak correction to the relevant Wilson coefficient, 
proportional to the difference $\Delta^{(2)}- \Delta_{eff}^{QED}(\mu)$,    
is free 
from infrared divergences, as the dependence on $q_t$ cancels against 
analogous terms in $\Delta^{(2)}$.
The Wilson coefficient, however,   
depends crucially on the scale $\mu$
at which the operator $\hat{Q}_{LL}$ is renormalized. 
 It is clear that our   analysis  also requires  the knowledge
of QED corrected matrix elements at the same scale $\mu$. At least 
in principle, they could be provided by dedicated lattice calculations,
but from a practical point of view it is likely that, since
the QED corrections turn out to be  very small, their impact would  be 
irrelevant. 

We now consider the evolution of the Wilson coefficient down to the bottom
mass scale. At lowest order in $\alpha_{QED}$, this amounts simply to 
setting $\mu=\mu_b$, with $\mu_b\approx m_b$ in \equ{deltaeff}.
By adapting the available QCD calculations
\cite{buras:90}  for the NLO 
anomalous dimension of the operator $\hat{Q}_{LL}$ to the case of QED,
the evolution of the Wilson coefficient down to $\mu_b$ 
could in principle be performed
consistently at NLO. However, 
as the QED coupling is very small and the QED logarithms have a negligible
effect on the result, this simple approach (setting $\mu=\mu_b$)
 seems  to us  sufficient. 
Indeed, we notice that the 
second term in the parenthesis on the r.h.s.\ of \equ{qedeff}, corresponding to the
leading logarithmic QED correction,  provides  only a 
$-0.3\%$ correction to the one-loop result  in the case
$\mu_b=4.8$\,GeV. 

Clearly, QCD corrections modify the evolution of the Wilson coefficient
 if the renormalization group equation 
 is solved keeping both QCD and QED effects into account.
Using $\alpha\ll \alpha_s$, 
the formalism is well known \cite{buras-harl:90,neubert}, 
and one finds that the second logarithmic term in \equ{qedeff} should be
multiplied by  the QCD screening factor $\eta'$,
given by
\be 
\eta'= \frac{4\pi}{\beta_0} \left(\frac1{\alpha_s(\mu_b)}
-\frac1{\alpha_s(\mt)}\right) 
 \left( \frac{\alpha_s(\mt)}{\alpha_s(\mu_b)}\right)^
{\frac{\gamma_0}{2\beta_0}}
 \left(\ln {\mu_b^2 \over M_t^2} \right)^{-1};
\ee
the rest of the Wilson coefficient undergoes the usual NLO
QCD evolution summarized by $\bar\eta_{2B}$.
Using $\beta_0=\frac{23}{3}$ and the QCD anomalous dimension $\gamma_0=4$ 
\cite{buchalla:96}, one obtains the numerical value $\eta' \approx 0.9$. 
This modest reduction of an already small effect
can be  approximately 
taken into account by noting that $\eta' \approx \bar \eta_{2B}$ and
 including the QED logarithm $\ln \mu_b^2 /M_t^2$ ({\em without} the
 screening factor) inside the curly parenthesis of \equ{effth}.
As a consequence of this discussion, we  identify the 
\ew correction $\delta S_{ew}$ of \equ{effth} 
with the Wilson coefficient evaluated at
 a scale $\mu_b$, which  for definiteness we set equal to $m_b$:
\be \label{matching}
\delta S_{ew}= \frac{g^2}{16\pi^2}
\left[\Delta^{(2)} - \Delta_{eff}^{QED}(\mu=m_b)\right].
\ee

\subsection{Calculation of the two-loop diagrams}
As far as  the two-loop diagrams contributing to $\Delta^{(2)}$ in
\equ{matching} are concerned, 
we need to consider all the topologies
shown in Fig.\,2,
\begin{figure}
\centerline{
\epsfysize=18truecm
\epsfxsize=18truecm
\epsffile{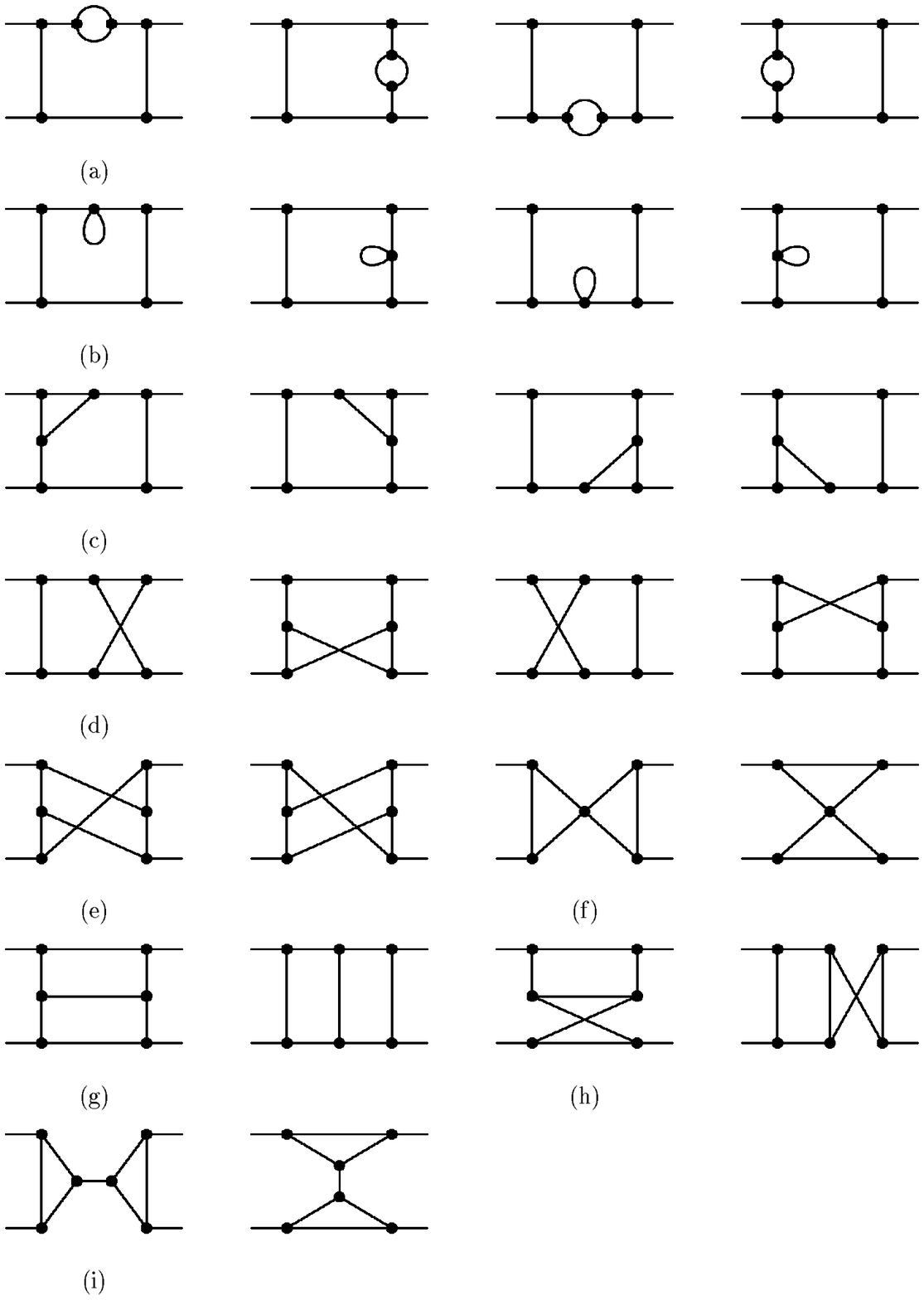}}
\caption{\sf The 26 two-loop topologies contributing to the  $B^0
\rightarrow \bar B^0$ transition at $O(g^6)$. 
}
\end{figure}
which correspond to a very large number of diagrams.
We have therefore decided to use the {\sc Mathematica} \cite{math} 
package {\it FeynArts 1.2}
\cite{feynarts} to generate automatically all the two-loop amplitudes.
Dirac algebra operations, reduction to scalar integrals, and substitution 
 of the scalar integrals have all 
been performed independently in two different 
ways  involving various combinations of the {\sc Mathematica} packages
{\it Tracer} \cite{tracer} and {\it ProcessDiagram} \cite{processdiag}, and of
{\sc Form} \cite{form}. The calculation has been performed in the 
't~Hooft-Feynman gauge, but we have checked the $\xi_Z$ independence.
As we neglect all external momenta, 
the complete result for the unrenormalized two-loop amplitudes can be 
written in terms of two-loop vacuum integrals, which admit a relatively compact
representation 
in terms of logarithms and dilogarithms of the internal masses
(see  e.\,g.\ \cite{davydychev}).  The final expression for the two-loop
diagrams, however, is very long, due to the presence of
four different heavy masses. Hence, we have decided to provide 
only approximate formulas that allow the reader to reproduce our numerical
results with high accuracy.
On the other hand, we feel that it may be more useful to explain 
in some detail the renormalization procedure that we have adopted. 
This is done in Sec.\,3, where we stress the importance of keeping it as 
simple as possible in
order to reduce the number of terms we have to deal with.
All partial and final results are available in full analytic
form from the authors.
The complete analytic result  has also allowed us to
perform various heavy mass expansions, which will be presented in Sec.\,4.. 

The computation of the contributions of individual two-loop diagrams to
the Wilson coefficient of the operator $\hat{Q}_{LL}$ can be considerably 
simplified by the implementation of a suitable projection procedure. 
In our case, such a projection can  be achieved by
forming appropriate traces 
(see e.\,g.\ \cite{herrlich:95} 
 for a nice explanation of the method):
 writing a generic two-loop amplitude as
\be
{\cal M} = \bar b \,D_1 \,d \otimes \bar b\, D_2 \,d\, ,
\ee
where $D_1$ and $D_2$ are   arbitrary Dirac structures 
with saturated Lorentz indices
(for instance $D_1 \otimes D_2 =\gamma^\nu \gamma^\lambda \gamma^\rho
(1-\gamma_5) \otimes \gamma_\nu \gamma_\lambda \gamma_\rho (1-\gamma_5)$),
 the projection on the
operator $Q_{LL}$
reads
\be \label{projection}
{\cal M} \rightarrow - {1 \over 256} \left ( 1+{3 \over 2} \epsilon
\right ) \, \mbox{tr} \left [ \gamma^\mu (1+\gamma_5) D_1
\gamma_\mu (1+\gamma_5) D_2 \right ] Q_{LL}\, ,
\ee
where terms up to $O(\epsilon)$ have been included.
 Furthermore, since $Q_{LL}$ represents 
the {\em only} operator contributing to the $B^0 - \bar B^0$
mixing at the order we are working at, 
\equ{projection} is in fact an identity.

A few  additional comments are now in order:
\begin{itemize}
\item We work in the framework of Naive Dimensional Regularization, a choice 
justified by the absence of any $\gamma_5$ ambiguity in the two-loop
graphs we have considered. In particular, no closed fermion loop
appears, apart from the ones in the self-energy insertions of the topologies in
Fig.\,2a, which pose no problem.

\item The number of diagrams is significantly reduced by the following
simple observation: for a given topology, one can consider subsets
of diagrams characterized by the number of light quark propagators they
contain and by the lines to which the light quarks are assigned. Since
all light quarks are treated as massless, diagrams within such 
subsets differ solely by the CKM factor, and can be grouped together, reducing
the total number of diagrams to be actually computed. 
Using the unitarity of the CKM matrix, the
overall CKM factor turns out to be always proportional to $\lambda_t^2$, as
expected from the above-mentioned {\it hard} GIM cancellation.

\item We perform the calculation in the \ew  scheme of Ref.\,\cite{msbar},  
which uses $\msbar$ couplings and on-shell masses as 
basic parameters. The transition to other popular schemes amounts 
to a finite renormalization of the \ew coupling and is therefore
 straightforward:  we discuss it in detail in the next section.
The residual $\msbar$ scale  and scheme dependence 
allows us to gauge the remaining ambiguity due to the truncation of the
perturbative series, as discussed in Sec.\,4.

\end{itemize}

\subsection{Double triangle diagrams}
As mentioned above, the full electroweak two-loop correction receives
 also a contribution $\Delta_{DT}$
from the
one-particle reducible topologies depicted in Fig.\ 2i. 
This contribution is finite but gauge dependent. 
The two penguin diagrams may be  
connected by $\gamma$, $Z$, $ H$, or $\phi^0$, but  
only the $\bar b dZ$ vertex  contributes to $\Delta_{DT}$
 for  vanishing external momenta and masses. Denoting this vertex by
$\Gamma^{\mu}_{\bar bdZ}$, one finds in the 't~Hooft-Feynman gauge
\cite{inami:81}
\be
i \Gamma^\mu_{\bar b dZ}= {g^3 \lambda_t \over 16 \pi^2 c}
 C(w_t) \, \bar b \gamma^\mu (1-\gamma_5) d,
\ee
with
\be
C(w_t)={6 w_t -1\over 8 w_t (w_t-1)} - {3+2 w_t \over 8 (w_t-1)^2} \ln
w_t .\label{Cfunc}
\ee
This result already includes the on-shell wave-function renormalization of
the external legs. For the ``double triangle" contribution $\Delta_{DT}$
we therefore obtain
\be
\Delta_{DT}= 16 \,C(w_t)^2.\label{deltaDT}
\ee
Numerically, $g^2/(16 \pi^2) \times (\Delta_{DT}/S_0) \approx+1.1\%$.


\section{Renormalization}
In renormalizing the two-loop amplitude our aim is to attain 
the maximal simplicity. We avoid all wave function renormalization 
of the internal lines, and choose a particularly simple procedure for 
the unphysical sector. 
All masses are defined on-shell and for the \ew coupling we use the $\msbar$
scheme \cite{msbar}, although we explain in detail the connection to other
schemes. 

In the following we give explicit expressions for the
various counterterms. They are written  in terms of logarithms and of 
a single 
function $B_0(x,y)$, which is defined through the one-loop 
integral 
 \bea
-{\rm Re}  \left(q^2 e^\gamma \right)^\epsilon 
\int \frac{d^n k}{\pi^{n/2}}
\frac{i}{[k^2-m_1^2][(k-q)^2-m_2^2]}=
\frac1{\epsilon} + B_0\left(\frac{m_1^2}{q^2},\frac{m_2^2}{q^2}\right)
 + O(\epsilon),
\label{B0}
\eea
whose analytic form is well known. The $O(\epsilon)$ part of the counterterms
is not needed. To the reader's convenience we report here 
the explicit expression
of $B_0$  for the three special cases that are needed in our calculation:
\bea  \label{b00a}
B_{0}(1,x)&=&2-{x \over 2} \, \ln x- \frac1{2} \,a(x), \non\\
B_{0}(0,x)&=&2-x \,\ln x-(1-x) \,\ln \vert 1-x \vert 
,\non\\
B_{0}(1,0)&=&2,\non
\eea
where the function $a(x)$ is given by 
\be
a(x)=\left \{ 
\begin{array}{cc}
2 \sqrt{4 \,x-x^2} \,\arctan \sqrt{4/x-1}, & 0<x\leq 4, \\
\sqrt{x^2 - 4\,x} \,\ln {1-\sqrt{1-4/x} \over 1+\sqrt{1-4/x}}, & x>4. \\
\end{array}
\right.
\ee

\subsection{Top Mass Counterterm}

The complete on-shell top mass counterterm 
is gauge-independent only after the inclusion of tadpoles, whose
 explicit expression  can be found, for instance, in \cite{ds}.
Here we report only the two-point function contribution 
to $\delta\mt$
in the 't~Hooft-Feynman
gauge. It can be written as
\bea
\frac{\delta \mt}{\mt}&=& \frac{g^2}{16\pi^2}\,
\left( \frac{\bar \mu^2}{ \mt^2}\right)^\eps
\left\{
-\frac3{8\eps} \left[ \frac1{w_t} +
\frac1{c^2}
\left(1-\frac{14}{9}s^2 \right)\right] +F_t + O(\eps)\right\},
\eea
where $F_t$ is given by
\bea
F_t&=&\frac1{w_t}\left[\frac{4-h_t}{16} a(h_t) +\frac{h_t-7}{8}
+\frac{6-h_t}{16} \, h_t\,\ln h_t\right]\non\\
&+& \frac1{4} +\frac{16}{9} s^2 -\frac1{8} (1+\frac1{w_t} -2 w_t)
B_0(0,w_t) -\frac{2w_t+1}{8} (1-\ln w_t) \non\\
&+&
\frac1{c^2}\left[ \left(\frac1{8} +
\left(\frac1{8} -\frac{s^2}{3} +\frac4{9} s^4\right)z_t\right)
\left( \ln z_t -1 + \frac1{z_t}\right)\right.\non\\
&+&\left.
\left(\frac{z_t-1}{8} + (2+z_t)\left(\frac49 s^2 -\frac13\right)s^2\right)
B_0(1,z_t)
+\frac18 +\frac{s^2}{3} -\frac49 s^4\right].\non
\eea
Replacing $M_t \rightarrow M_t - \delta M_t$
in the LO result (\ref{oneloopres}), one finds the following top counterterm
contribution to be added to $\Delta_{2loop}^{unren}$:
\be
\Delta^c_{M_t}= 2 \,w_t \left ( {\bar \mu^2 \over M_t^2}
\right )^\epsilon 
\left \{ -{3 \over 8 \epsilon} \left[{1 \over w_t}+{1 \over c^2} \left
( 1-{14 \over 9} s^2 \right ) \right ] \tilde S'+ S_0' \,F_t \right
\},\label{deltacmt}
\ee
with $\tilde S'\equiv\partial \tilde S(w_t,\bar \mu^2/\mw^2) / \partial
w_t$,
 $S_0' \equiv
\partial S_0/ \partial w_t$, the function 
$S_0(w_t)$ as given in \equ{S0function}, and $\tilde S$ defined by
\bea \label{Stildefunction}
\tilde S(w_t,\bar \mu^2/\mw^2)= S(w_t,w_t\, \bar \mu^2/\mw^2),\non
\eea
where $S(w_t,\bar \mu^2/M_t^2)$ has been given in \equ{Sfunction}.

The top quark mass is therefore renormalized 
{\it on-shell} for what concerns 
electroweak effects\footnote{A  $\msbar$ top mass renormalization in the 
\ew sector would make the renormalized top mass dependent on 
the precise value of the Higgs mass, which is unknown; 
we do not consider it here.}, 
while $\bar\eta_{2B}\approx 0.85$ in \equ{effth} implies the 
use of a $\msbar$ mass $\mtbar(\mt)$ as far as  QCD effects are considered, in
accordance to the standard convention \cite{buchalla:96}.
Using the LO QCD relation between {\it on-shell} and the $\msbar$ mass
and the pole mass value $\mt=174$\,GeV \cite{LEP}, we find 
$\mtbar(\mt)=166$\,GeV, which will be our input in the following and will be
denoted for simplicity just by $\mtbar$.

\subsection{Renormalization of the unphysical sector}

Before considering the renormalization of the 
$W$ mass it is necessary to explain the procedure we have followed 
for the unphysical scalars.
As is well known, the renormalization of the unphysical sector
is not independent from the way the physical sector is treated.
Indeed, 
the renormalization procedure must respect the Slavnov-Taylor Identities
(STI)
which are induced by the local gauge invariance of the original Lagrangian
before spontaneous symmetry breaking.
According to the organization of the calculation,
it is possible to use different procedures that respect the STI's
and are particularly convenient in order, for example, 
to minimize the number of counterterms to be
considered. Of course, physical amplitudes are independent of the 
chosen procedure, and this can be used as an
additional check of the calculation.
For the problem at hand, the discussion can be kept at the one-loop level.

If we split the unrenormalized $W$ polarization tensor  into 
transverse and longitudinal parts
\be
\Pi^{\mu\nu}_\smallw(q)= \left(g^{\mu\nu} -\frac{q^\mu q^\nu}{q^2}\right) 
A_{\smallw\smallw}(q^2)
+ \frac{q^\mu q^\nu}{q^2} \Pi^{{\scriptscriptstyle L}}_{\smallw\smallw}(q^2),
\ee
and we denote by $\Pi_{\smallw\phi}$ 
the two-point function for the  mixing between the $W$ and its
pseudo-Goldstone boson, $\phi$, and  by $\Pi_{\phi\phi}$ the self-energy of the
latter, 
we obtain the following STI in a general $R_\xi$ gauge 
(see for ex. \cite{aoki:82,boehm:86})
\be
q^2\left( \Pi^{{\scriptscriptstyle L}}_{\smallw\smallw}(q^2) + 2 \mw 
\Pi_{\smallw\phi}
(q^2)\right) + \mw^2 \,\Pi_{\phi\phi}(q^2) + \mw^2 \,T=0\, ,
\label{ident}
\ee
where $T$ represents the tadpole  contribution.
In particular, at $q^2=0$, the first two terms vanish, and
the STI implies the cancelation between 
$\Pi_{\phi\phi}(0)$ and the tadpole contribution. This uncovers 
the connection between  the renormalization of the Goldstone boson mass 
and  the one of the tadpole. 
As anticipated, the counterterm contributions to the various terms
in \equ{ident} must also respect the STI.
In practice, the usual tadpole renormalization that minimizes the effective 
potential    and consists in removing all tadpole graphs implies the 
subtraction of $\Pi_{\phi\phi}(0)$ from the two-point function of the
pseudo-Goldstone boson\footnote{This point is nicely explained in Taylor's
book \cite{taylor}, see section 14.6.}.
As mentioned above, we adopt the physical mass to define the masses of the 
vector bosons. 
It is therefore convenient to renormalize the longitudinal 
component of the two-point function of the $W$ in the same way as the
transverse, using $\delta \mw^2=
{\rm Re}\,A_{\smallw\smallw}(\mw^2)$. 
For the other two-point functions different options are possible, which all
respect the STI, and are equivalent at the level of physical amplitudes.
They correspond to different ways of renormalizing the gauge-fixing parameters.

One possibility consists in assigning  no counterterm 
to the $W-\phi$ transition; in the 't~Hooft-Feynman gauge 
this corresponds to renormalizing 
the masses of the 
vector boson and of the associated scalar boson in the same way
({\it bare gauge fixing}).
Of course, the mass of the scalar boson will still 
need a supplementary subtraction at $q^2=0$, corresponding 
to the tadpole contribution.
This choice  clearly verifies \equ{ident} at $q^2=\mw^2$, leaving 
room for a further arbitrary  wave-function renormalization, which we avoid
altogether, as the $W$ boson appears only inside the loops.
Because of its simplicity, this is our preferred option:
it amounts to fixing $\delta M^2_\phi=\delta \mw^2 + 
T$  and is the closest to the naive parameter
renormalization.

Another possibility, for example, would imply the subtraction of 
the first two terms of the  Taylor expansion around $q^2=\mw^2$
of the individual two-point functions in the external momentum, 
and it would obviously respect the identities, as the 
unrenormalized self-energies do. A counterterm for the $W-\phi$
transition is now involved.
We have explicitly verified that all the renormalization options that satisfy 
the STI are equivalent 
at the level of physical amplitudes.

Finally, we recall  that the renormalization of the three point  
function $\bar d u\phi$ is fixed by the Ward identity that links the 
Yukawa coupling  counterterm 
to the gauge coupling and fermion mass renormalization.

\subsection{W mass  counterterm}
Following the strategy outlined in the preceding subsection, we use the same
coun\-ter\-term to renormalize the $W$ and the pseudo-Goldstone boson mass terms.
In the 't~Hooft-Feynman gauge, one has for the two-point 
contributions to the $W$ mass counterterm
(see e.\,g.\  \cite{marciano:80}):
\be \label{zmw}
\frac{\delta \mw^2}{\mw^2}=
{\rm Re}A_{\smallw\smallw}(\mw^2)=
{g^2 \over 16 \pi^2} \left ( {\bar \mu^2 \over M_t^2}
\right )^\epsilon \left \{ {1 \over \epsilon} \left [{3 \over 2 w_t}-{1
\over c^2}+{7 \over 6} \right ]  +F_\smallw
+O(\eps) \right \}, 
\ee
where the finite part $F_W$ is given by
\bea
F_\smallw&=& \left ( {1 \over 2 w_t^2}+{1 \over 2 w_t}-1 \right ) 
B_{0} \left ( 0,{1 \over w_t} \right )
- \left ( {h_t^2 \over 12 w_t^2}-{h_t \over 3 w_t}+1 \right )
B_{0} \left (1,{h_t \over w_t} \right ) \nonumber \\
&& + \left ( 4 c^2+{17 \over 3} -{4 \over 3 c^2}-{1 \over 12 c^4}
\right )
B_{0} \left ( 1,{z_t \over w_t} \right )- \left ( 4-{5 \over 6 c^2}-{1 \over 6 c^4}  \right ) \ln c  \nonumber \\
&&- \left [ {1 \over w_t^2} \left ( {1 \over 2}-{h_t^2 \over 12}
\right ) +{1 \over w_t} \left ({1 \over 2}+{h_t \over 4} \right ) + {7
\over 6} -{1 \over c^2} \right ] \ln w_t  - \left ( {h_t^2 \over 12 w_t^2}-{h_t \over 4
w_t} \right ) \ln h_t \nonumber \\
&&
-8 c^2-{1 \over w_t^2} \left ({1 \over 2}-{h_t^2 \over 12} \right )
-{h_t \over 6w_t}  
+{53 \over 18}+{1 \over 2 c^2}+{1 \over 12 c^4}. \non
\eea
Replacing $\mw \rightarrow \mw - \delta \mw$
in the LO result \equ{oneloopres}, one obtains the $W$ mass counterterm
contribution 
\be
\Delta_{\mw}^c= \left ( {\bar \mu^2 \over M_t^2}
\right )^\epsilon \left \{ {1 \over \epsilon} \left [ {3 \over 2
w_t}-{1 \over c^2}+{7 \over 6} \right ] (S-w_t \,S')+(S_0-w_t \,S_0')
\,F_\smallw 
\right \}.\label{deltacmw}
\ee
Here again the notations $S'\equiv\partial S(w_t,\bar \mu^2/\mw^2) / \partial
w_t$ and 
 $S_0' \equiv\partial S_0 / \partial w_t$ are understood, 
and the functions $S_0(w_t)$, $S(w_t,\bar \mu^2/M_t^2)$ are given in
Eqs.\ (\ref{S0function}) and (\ref{Sfunction}).
We  treat the $W$ mass as a fundamental 
input parameter, whose value is taken directly from the experiment:
$\mw=80.385\pm 0.065$ \cite{LEP}. Although the determination of $\mw$
from $\mz$, $G_\mu$, and $\alpha$ is presently more precise than the
experimental one after including all theoretical uncertainties \cite{DGPS}, 
this is certainly sufficient for our purposes.

\subsection{Tadpole contribution}
As mentioned above, the renormalization of the pseudo-Goldstone boson mass
term  can be performed (in the 't~Hooft-Feynman gauge) in the same way as 
the one of the $W$ mass, apart from an additional tadpole contribution, whose
physical origin has been already explained.
In practice, this means that the one-loop diagrams containing 
scalars induce  a further counterterm contribution that can be 
easily calculated from the complete one-loop tadpole (see for instance 
Eq.\,(12) of \cite{ds}) and the one-loop amplitudes involving only scalars.
Here we just give the final result for this additional 
contribution, again  in the 't~Hooft-Feynman gauge:
\be
\Delta_{Tad}^c=\left ( {\bar \mu^2 \over M_t^2} \right )^\epsilon
\left [ E_{Tad} \left({T_0 \over \epsilon} +T_1\right)
+F_{Tad} \,T_0  +O(\eps) \right] \, ,
\label{deltactad}
\ee
with 
\bea
T_0 &=&{-1+19 w_t \over 4 (w_t-1)^3}+ {1-6 w_t-4 w_t^2 \over 2
(w_t-1)^4} \ln w_t, \non\\
T_1 &=&{-13+43 w_t \over 8 (w_t-1)^3} + {4-19 w_t^2 \over 4 (w_t-1)^4}
\ln w_t- {1- 6 w_t-4 w_t^2 \over 4 (w_t-1)^4} \ln^2 w_t+T_0 \ln {\bar
\mu^2 \over M_t^2},\non\\
E_{Tad}&=& -{3 \over w_t^2}+{3 h_t^2 \over 8 w_t^2}+{h_t \over 4 w_t}+{h_t
\over 8 c^2 w_t}+{3 \over 2}+{3 \over 4 c^4}\non ,\\
F_{Tad}&=&E_{Tad}-1-\frac1{2c^4}
-{3 h_t^2 \over 8 w_t^2} \ln h_t 
- \left ({3 \over 2}+{h_t \over 4 w_t}
\right) \ln w_t-\left ( {3 \over 4 c^4}
+{h_t \over 8 c^2 w_t} \right ) \ln z_t.\non
\eea

\subsection{Coupling counterterm}
The SU(2) coupling counterterm in  the $\overline{MS}$ scheme is 
(up to  $\ln 4\pi$ and $\gamma_E$ constants)
\bea
\frac{\delta \hat{g}}{\hat{g}}= {\hat{g}^2 \over 16 \pi^2} 
{19 \over 12 \epsilon}\,. \nonumber
\eea
Performing the renormalization
$g_0= \hat{g} -\delta \hat{g}$, we obtain the following contribution to
$\Delta^{(2)}$:
\be
\Delta_g^c= - \frac{19}{3\eps} \, S(w_t,\mu^2/\mt^2).
\label{deltacg}
\ee
In the $\msbar$ scheme $\hat{g}(\mz)$ can be calculated from $\hat{\alpha}
(\mz)=(128.1)^{-1}$  and $\hat{s}^2(\mz)\approx \sin^2\theta_{eff}^{lept}
- 1\times 10^{-4}=0.23145$ \cite{DGS}, obtaining 
$\hat{g}^2(\mz)=0.423842$; this is the value that will be used in the
following numerical calculations. 

Let us now consider what happens in different renormalization schemes.
In any scheme $X$ but $\msbar$ 
the counterterm would have a finite part $F_g^X$, 
namely
\bea
\frac{\delta g^X}{g}= {g^2 \over 16 \pi^2} 
{19 \over 12 \epsilon} + F_g^X ,\nonumber
\eea
so that  the difference between the result in a scheme $X$ and the
result in the $\msbar$ scheme at the scale $\mu_g=\mz$ is just
\be
\delta S_X= -4 \, S_0(w_t) \,F_g^X,
\ee
which should be added to  $\delta S_{ew}^{\msbar}$,
also evaluated at $\mu_g=\mz$.
It is well-known  that
the use of the electromagnetic fine structure constant 
$\alpha(0)= 1/137.036$  to normalize the \ew coupling 
would introduce large and
uncompensated mass singularities \cite{si80} in the two-loop result. 
Therefore,  the coupling
$g$ must be normalized at short distances. This is the case, for instance,
if we use  $G_\mu$ and $\mw$:
$g^2= 4\,\sqrt{2} \,G_\mu \,\mw^2$.
Then we find
\be
F_g^{G_\mu}=\frac12 \left(\Delta\hat{r}_\smallw+ \left.\frac{2\delta e}{e}
\right|_{\msbar}\right)
\label{fgmu},
\ee
where $\Delta\hat{r}_\smallw$ is a function of the top,  Higgs, $W$ and $Z$
masses, as well as of the couplings, which summarizes the \ew corrections to
the muon decay in the $\msbar$ scheme 
and is given explicitly in \cite{msbar}. The second term in
parenthesis, instead, represents the finite part of the conventional 
electric charge counterterm \cite{si80}, evaluated at the $\msbar$ scale 
$\mu_g$.
$F_g^{G_\mu}$ depends mildly (logarithmically) on $\mh$ and $\mt$; numerically
it is very small, as can be seen comparing  $g^2$ obtained from
$G_\mu \mw^2$ 
and the $\msbar$ value of $\hat{g}^2(\mz)$: they differ by
half a percent.

We will consider later also another possibility
 to define $g$,   namely through the relation
$g^2 = 4\pi \,\hat\alpha(\mz)/s_\smallw^2$.
Here we have adopted the notation   $s^2_\smallw=
1- \mw^2/\mz^2$ for the {\it on shell} 
definition of the weak mixing angle \cite{si80}
and $\hat\alpha(\mz)$ stands for the 
$\msbar$ running electromagnetic coupling at the scale $\mz$.\footnote{To keep
the notation simple,  we
use here the $\msbar$ running electromagnetic coupling $\hat\alpha(\mz)\simeq
(128.1)^{-1}$ instead of the conventional $\alpha(\mz)\simeq
(128.9)^{-1}$ \cite{Jeger}, which is defined by subtraction of 
the renormalized light fermion photon correlators only, and is more 
generally used in the context of the {\it on-shell} scheme. 
The difference is not negligible, and must be taken into account in the
two-loop corrections.}
In this case
\be
F_g^{OS}= \frac12 \,
\frac{c_\smallw^2}{s_\smallw^2} \,\,\Delta\hat{\rho}\, ,
\ee
where $\Delta\hat{\rho}$ is another function of the heavy masses given in 
Ref.\cite{msbar}. Unlike \equ{fgmu}, however, $F_g^{OS}$ is very sensitive 
to $\mt$, due to its quadratic dependence enhanced by the factor 
$ c_\smallw^2/s_\smallw^2$. Indeed, the $\msbar$ and {\it on-shell} definitions
of $\sin^2\theta_\smallw$, and so the corresponding $g^2$,
 differ by almost 4\%\,!
We can therefore anticipate that 
this normalization choice will introduce large two-loop
corrections when compared to the $\msbar$ scheme. 
Moreover, as the origin of the
large counterterm contributions has nothing to do with the process at hand, 
we can also expect this choice to give unnaturally large two-loop
corrections.

\subsection{Wave function counterterm}
The left-handed quark fields of the down type (i.\,e. the external
fields) are renormalized according to
\be
{(d^L_i)}^0=\left [ (Z^L_d)^{1/2} \right ]_{ij} d^L_j,
\ee
and the (matrix-valued) wave-function renormalization constant
$Z^L_d=1+ \delta Z^L_d$
is determined as described by Aoki {\em et al.}\
\cite{aoki:82}. Denoting the sum of all one-loop diagrams contributing
to the transition $d_i \rightarrow d_j$ with external momentum $p$ by
 $i \Sigma_{ij}(p)$, and writing
($L,R=(1 \mp \gamma_5)/2$)
\be
\Sigma_{ij}(p)= \Sigma^L_{ij}(p^2) p \fsl L + \Sigma^R_{ij}(p^2) p
\fsl
R+ \Sigma^S_{ij}(p^2) (m_i
L+m_j R),
\ee
this prescription implies in the limiting case of vanishing masses of the down-type quarks the customary relation
\be \label{zsigmarelation}
(\delta Z^L_d)_{ij}= - \Sigma^L_{ij}(0).
\ee
In this equation we have considered only the hermitian part of the
  wave function renormalization constant.
For what  concerns the antihermitian part of $\delta Z_d^L$, 
it is not defined in the limit of
vanishing  quark masses that we have adopted from the beginning. 
One can use the renormalization prescription of
Denner and Sack \cite{denner:90} for the CKM matrix elements and remove 
it completely. In the case of the up quarks, the top mass cannot be
set to zero, but the antihermitian part of the wave function renormalization
vanishes because of the GIM mechanism. 

By calculating the relevant diagrams, we obtain in this way
\be \label{wfrenom}
(\delta Z^L_d)_{ij}={g^2 \over 16 \pi^2}  \left ({\bar \mu^2 \over m_t^2}
\right ) ^\epsilon \biggl ( \delta_{ij} A+V_{ti} V_{tj}^\ast B \biggr )
,\ee
where 
\bea
A&=& {1 \over c^2} \left [ -{1 \over \epsilon} \left ({1\over 36}+
{13 \over 18} c^2 \right ) + {1 \over 72} +{1 \over 12}
c^2+{5 \over 18} c^4+ \left ({1 \over 36}+{1 \over 9} c^2+{1 \over 9}
c^4 \right ) \ln z_t  \right ]+ {1 \over 2}  \ln w_t \nonumber \\
&&+{1\over 9} s^2 \ln q_t ,\\
B&=& - {1 \over \epsilon} {1 \over 4 w_t} - {3 (1+w_t) \over 8 w_t
(1-w_t)}+{w_t-4 \over 4 (1-w_t)^2} \ln w_t .
\eea

The corresponding counterterm contribution $\Delta^c_\psi$ is obtained
as follows: the amputated amplitude ${\cal M}^{(k)}_{1loop}$ for the transition
$\bar k+ d \rightarrow b+\bar d$ ($k=d,s,b$)  
is in the notation of
 Sec.\ \ref{sec:oneloop} given by
\be
{\cal M}_{1loop}^{(k)}={-i \over 16 \pi^2} {g^4 \over 8 \mw^2}
\sum_{i,j} \lambda_i^{(k)} \lambda_j \, {\cal
S}^{(i,j)} \, Q_{LL},
\ee
with $\lambda_i^{(k)}=V_{ik}^* V_{id}$. Using the unitarity of the CKM
matrix, this simplifies to
\be \label{mk1loop}
{\cal M}_{1loop}^{(k)}={-i \over 16 \pi^2} {g^4 \over 8 \mw^2} \left
\{ \lambda _t^{(k)} \lambda_t \left [{\cal S}^{(t,t)}-2 {\cal
S}^{(t,c)}+{\cal S}^{(c,c)} \right ] + \delta_{kd} \lambda_t \left
[{\cal S}^{(t,c)}-{\cal S}^{(c,c)} \right ] \right \}.
\ee
The second term contributes only for $k \equiv d$, and it originates
in the unitarity relation
$\lambda_t^{(k)}+\lambda_c^{(k)}+\lambda_u^{(k)}=\delta_{kd}$. The
counterterm contribution due to the wave function renormalization of
the external $\bar b$-field is then
\be
\Delta^{c,1}_{\psi}= {1 \over 2} \sum_{k} \left [ \delta Z^L_d \right
]^*_{bk} {\cal M}_{1loop}^{(k)}.
\ee
Combining (\ref{wfrenom}) and (\ref{mk1loop}), one thus obtains
\be 
\Delta^{c,1}_\psi = {1 \over 2}  \left ({\bar \mu^2 \over m_t^2}
\right ) ^\epsilon \left [ (A+B) \, S(w_t,\bar \mu^2/M_t^2) +B \,
U(w_t,\bar\mu^2/M_t^2),
 \right ]
\ee
with $S(w_t,\bar \mu^2/M_t^2)$ as given in (\ref{Sfunction}), and
\bea
U(w_t,\bar \mu^2/M_t^2) &\equiv &{\cal S}^{(t,c)}-{\cal S}^{(c,c)} =
\left({1 \over w_t-1}- {w_t \ln w_t \over (w_t-1)^2} \right)
\left(1+ \eps \, \ln {\bar \mu^2 \over M_t^2}\right)\non\\
&+&  \eps\left(- {1 \over 2 (w_t-1)}+ {2-w_t \over 2 (w_t-1)^2}
\ln w_t+ {w_t \ln^2 w_t \over 2 (w_t-1)^2}\right) +O(\eps^2)\non
\eea
An analogous reasoning holds for the wave function renormalization of
the other three external fields, and the final counterterm
contribution is simply
\be
\Delta^c_\psi=4 \,\Delta^{c,1}_\psi.\label{deltacpsi}
\ee


\section{Results and discussion}
Combining the counterterm contributions of
Eqs.\,(\ref{deltacmt}), (\ref{deltacmw}--\ref{deltacg}), (\ref{deltacpsi})
 with \equ{deltaDT} and with the unrenormalized two-loop
contributions yields our final result in the $\msbar $ scheme.
The size of the effect can be seen in Fig.\,3, where the  following
\begin{figure}
\centerline{
\epsfysize=13truecm
\epsfxsize=8truecm
\rotate[r]{\epsffile{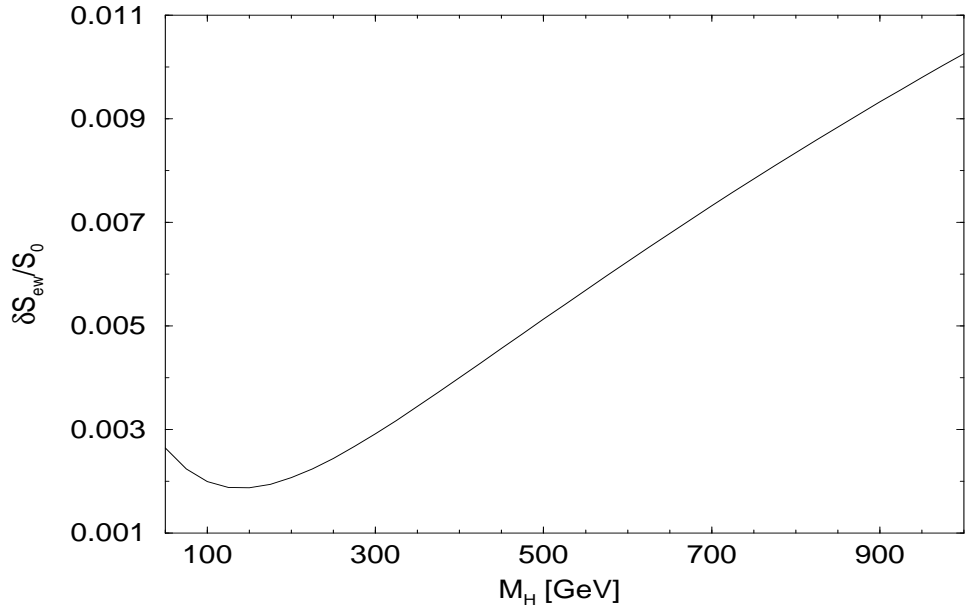}}}
\caption{\sf The \ew correction $\delta S_{ew}$ in the $\msbar$ scheme, 
normalized to the LO result, as a function of the Higgs
mass. Inputs as in the text. 
}
\end{figure}
input values have been used \cite{LEP}:
 $\mw=80.385$\,GeV, $\mu_g=\mz=91.187$\,GeV, $\mtbar=166$\,GeV, 
$\mu_b=4.8$\,GeV, $\hat{s}^2(\mz)=0.23145$. These are the values we will use
throughout this section.
The \ew correction is less than  1\% of the LO result
 for any value of the Higgs mass 
below 1\,TeV,
 and is particularly small for a light Higgs, namely in the region 
 preferred by recent global fits \cite{LEP,DGPS}.
The logarithmic asymptotic dependence for large Higgs masses is already evident
around 1 TeV. In Sec.\,4.3 we will present a simple approximate formula that
reproduces the complete result very accurately.
Before, however, we consider the heavy mass expansions 
of the full result and study the residual scheme and scale dependence.

\subsection{Heavy mass expansions}
We first consider the heavy Higgs expansion. In the case of a very heavy
Higgs boson, $\mh\gg\mw ,\mt$, we have verified that the leading behaviour is
logarithmic in $\mh$. In particular,  the $\msbar$ scheme result reads
\bea
\delta S_{ew}^{hh}=
\frac{g^2}{16\pi^2} \frac{1}{w_t^2}\left[\frac{ \ln^2  h_t}{32} 
+ \left(  {\frac{8w_t^3 - 27w_t^2+ 30w_t-5}{48{{\left( w_t-1 \right) }^2}}
     } + \frac{w_t^2 \,( 3w_t-5 ) \,
       \ln w_t}{16\left(w_t-1 \right)^3}
\right)\ln h_t + \ldots
\right]\label{leadh}
\eea
where the ellipses stand for constants or terms suppressed by inverse powers of
the Higgs mass. 
Of course, in another scheme $X$ the Higgs dependence contained in 
$\delta S_X$ will change the $\ln h_t$ term in the above expression. 
The absence of a  leading quadratic 
Higgs behaviour in  the complete result can be explained
in the context of an effective theory approach where the Higgs boson is
integrated out \cite{longhitano}. 

If we are interested in  the leading heavy-top
corrections, $O(g^2 \,\mt^4/\mw^4)$, we may work in the framework of a
Yukawa Lagrangian where the heavy fermions couple only to the Higgs boson and
to the longitudinal components of the gauge bosons, a situation which
corresponds to the {\it gaugeless limit} of the SM \cite{barb}.
In the covariant gauges, it suffices to  consider only the
diagrams that involve exchanges of scalar bosons. Even in the case of a heavy
Higgs boson, in which the  scalar coupling $\lambda \sim g^2\,\mh^2/\mw^2$ 
cannot be neglected 
with respect to the Yukawa coupling of the top, only a few tens of two-loop
diagrams contribute. Notice that in this framework the mass of the Higgs boson
is completely arbitrary. 

Computing the various counterterms from  the {\em scalar}
one-loop diagrams only (which corresponds to the heavy top limit of our
renormalization constants),  adding these to the sum of the {\em
scalar} two-loop diagrams,  and expanding in small $w_t$, we get
\bea
\delta S^{lead}_{ew}&=& {g^2 \over 16 \pi^2} {1 \over w_t^2} \Biggl [
{-44 +112 h_t-66 h_t^2+14 h_t^3-h_t^4 \over 64} \, \phi \left
({h_t \over 4} \right ) 
 + {\frac{52 - 16\,h_t - 9\,h_t^2}{32}}
\nonumber\\
&&
  -\frac{ \left(h_t-1 \right)^2 \left( 4 - 14\,h_t + h_t^2 \right)   
      }{16}
\, \mbox{Li}_2(1-h_t)
-{22 h_t^2-20 h_t^3+h_t^4 \over 64} \, \ln^2 h_t \nonumber \\
&&
 - \frac{h_t \left( 14 + 9\,h_t \right)  }{32} \ln h_t
- {8 +22 h_t^2- 20 h_t^3+h_t^4
\over 192} \, \pi^2
+\frac{h_t-4}{32} \, a(h_t) \Biggr ] .
\label{lead}
\eea
Here we have indicated the dilogarithmic function as 
${\rm{Li_2}} (x) = - \int_0^x dt \, {\ln (1-t) \over t} $, 
and introduced 
\be
\phi(z) =\left\{
\begin{array}{cc}
              4 \sqrt{{z \over 1-z}} ~\mbox{Cl}_2 ( 2 \arcsin \sqrt z ), &
  0 < z \leq 1,\\
  { 1 \over \lambda} \left[ - 4 {\rm Li_2} ({1-\lambda \over 2}) +
       2 \ln^2 ({1-\lambda \over 2}) - \ln^2 (4z) +\pi^2/3 \right] 
       ,&   z >1 ,\\
\end{array}
              \label{phi}\right.
\ee 
where $\lambda = \sqrt{1 - {1 \over z}}$ and $\mbox{Cl}_2(x)= {\rm Im} 
\,{\rm Li_2} (e^{ix})$ is the Clausen function.
It is a welcome check of our result that  \equ{lead}  coincides with 
the limit $w_t\rightarrow 0$ of the complete result, which is scheme and scale
independent. The limit for $\mh\to 0$ of the square parenthesis in
\equ{lead} is $\frac{13}{8}- \frac{\pi^2}{12}$. One can also verify that the
heavy Higgs limit of \equ{lead} coincides with the heavy top limit of
\equ{leadh}.

Numerically, $\delta S_{ew}^{lead}/S_0=1.12\%$ (for $\mh=100 \, \mbox{GeV}$),
which should be compared to 0.18\% obtained using the full result.
Clearly, the leading term of a heavy top expansion does not approximate 
well the complete result. This is not surprising since, for the measured
value of the top quark mass,
the heavy top expansion of the LO result,
\be
S_0(w_t)=  {\frac{1}{4\,w_t}}  -{\frac{9}{4}} -{\frac{3}{2}}\,\ln w_t
-w_t \left({\frac{15}{4}} + {\frac{9}{2}}\ln w_t\right) + O(w_t^2)
\ee
must be pushed up to the fourth term 
in order to reach a 10\% accuracy.
An example of slow convergence of the heavy top expansion can also be found
in the $O(\alpha\alpha_s)$ corrections to the $Z\to \bar b b$ decay amplitude
\cite{stein:98}, although in that case the leading top contribution
dominates, because of large cancellations among the subleading terms.

It is therefore interesting to investigate further the convergence of the
heavy top expansion in a case where the complete two-loop contributions are
available in analytical form. We have expanded our complete $\msbar$ result 
in inverse powers of the heavy top mass, up to the third term, namely 
excluding only contributions which are formally $O(\mw/\mt)$.
With respect to the Higgs boson mass, we either consider it much smaller than
the top mass, or comparable, i.e. heavy compared to $\mw$.
Consequently, we obtain two expansions, one valid for small $\mh$, the other
for large $\mh$. In Fig.~4, we compare them with the leading top result 
of \equ{lead} and with the complete result; we display the various 
results as functions of $\mh$ for the two cases $\mtbar=166$\,GeV and 
$\mtbar=350$\,GeV, respectively.
\begin{figure}
\centerline{
\epsfysize=8truecm
\epsfxsize=6truecm
\rotate[r]{\epsfbox{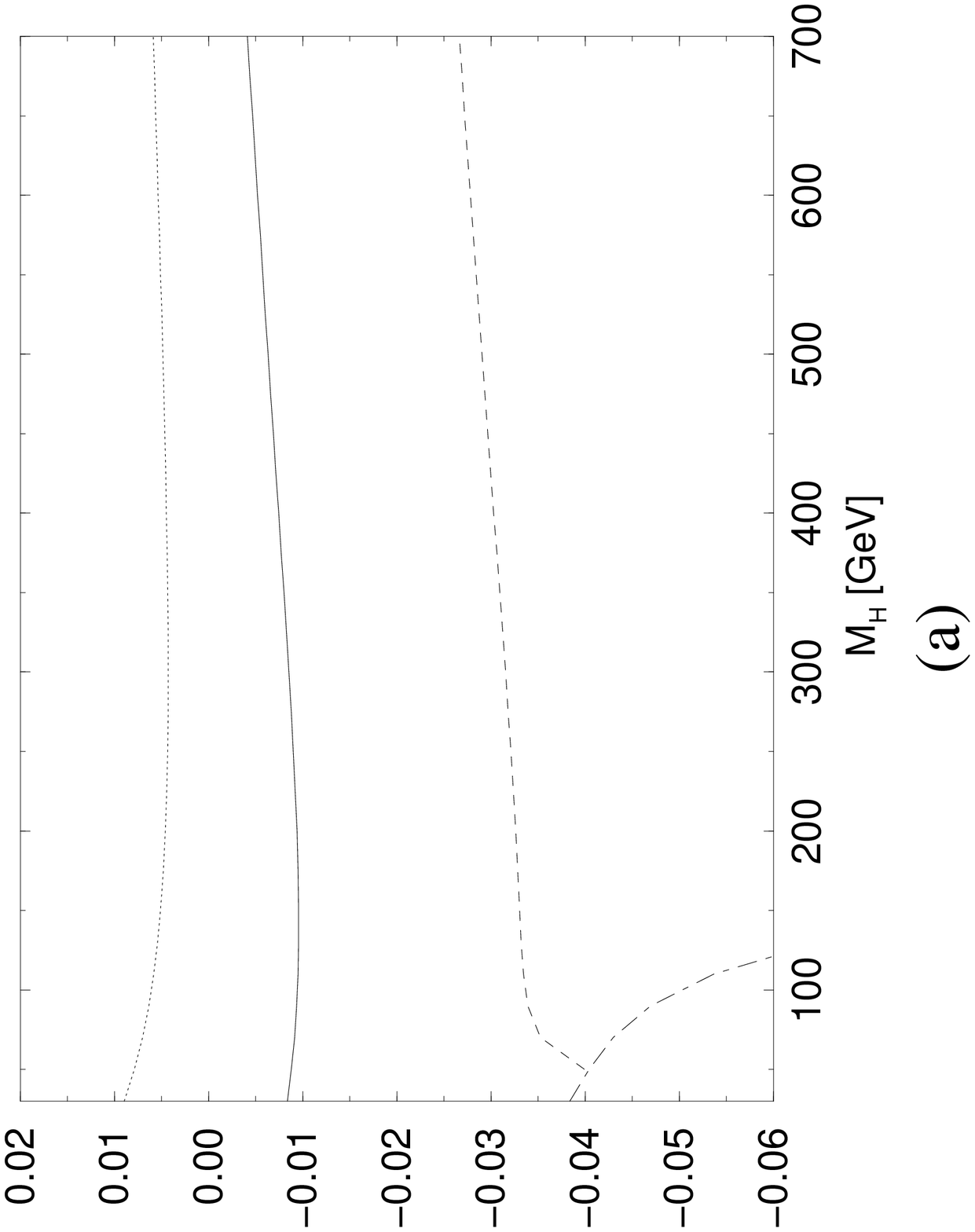}}
\epsfysize=8truecm
\epsfxsize=6truecm
\rotate[r]{\epsfbox{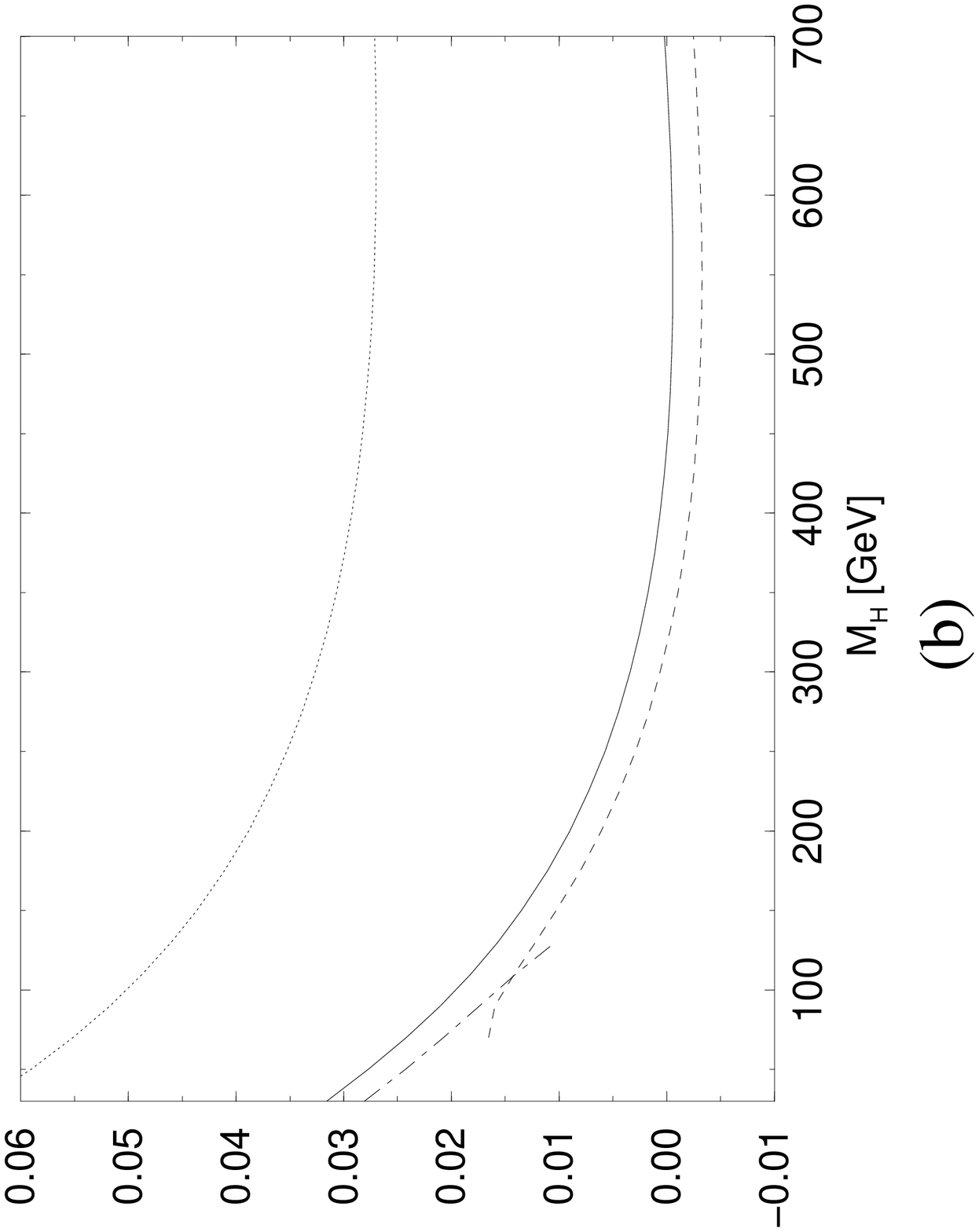}}}
\caption{\sf (a) Renormalized box contributions to $\delta S_{ew}$ 
in the $\msbar$ scheme, 
normalized to the LO result, as a function of the Higgs
mass. The solid line represents the complete result, the dotted line the
leading top approximation, and the dashed (dashed-dotted) line the 
heavy (light) Higgs top expansions up to $O(1/\mt)$. Inputs as in the
text. (b) The same, but for $\bar M_t=350$\,GeV, demonstrating the much
better convergence of the heavy top expansions in this case.
}
\end{figure}
We show only the contributions of the renormalized box diagrams, i.e.
we leave aside the double triangle contribution $\Delta_{DT}$; the latter
does not depend on $\mh$ and its top expansion converges also very slowly,
as can be seen from its explicit expression of \equ{deltaDT}.
We observe that in both cases the light and heavy Higgs expansions match 
quite well around $\mh\approx 60$ and 100\,GeV
 (similar to what happens in \cite{DGV}),
but start to converge to the complete result only for a very heavy top, with 
$\mtbar=350$\,GeV. 
For even heavier top masses, the expansion up to $O(1/\mt)$ terms
provides an  excellent approximation.

The slow convergence of the heavy top expansion of the two-loop contributions
may be  related to the slow  convergence of the expansion of the LO result.
It  is therefore quite different from  what happens in 
the case of the calculation of precision observables \cite{DGV,DGS,cks}, where
the top contributions  originate from two-point functions.
The heavy top mass expansion for the latter 
seems to work remarkably well up to two-loop \cite{weiglein} 
and has been checked even at 
the three-loop level in the case of mixed $O(\alpha\alpha_s^2)$
corrections \cite{cks}.

\subsection{Scale and scheme dependence}
In the $\msbar$ scheme, we have so far fixed $\mu_g=\mz$.  However,
by varying $\mu_g$ around $\mw$, we 
can compare the dependence on the  renormalization scale 
$\mu_g$ of the LO result with the  one of the sum of one and two-loop
results. We recall that the LO result is expected to exhibit a 
strong scale dependence because of the factor $\hat{g}^4(\mu_g)$. Our
two-loop correction cancels the leading logarithmic dependence on $\mu_g$, and
the remaining much weaker scale dependence can be interpreted as an 
indication of the 
importance of higher order effects.
This is illustrated in Fig.\,5, where we show that in varying the scale $\mu_g$
between $\mw/2 $ and $2\mw$  the
combination $\hat{g}^4 \, S_0(w_t)$ varies by  about 5\%, while 
the combination  $\hat{g}^4 \, (S_0(w_t)+\delta S_{ew})$ 
undergoes a maximum variation of less than 0.1\%.
\begin{figure}
\centerline{
\epsfysize=13truecm
\epsfxsize=8truecm
\rotate[r]{\epsffile{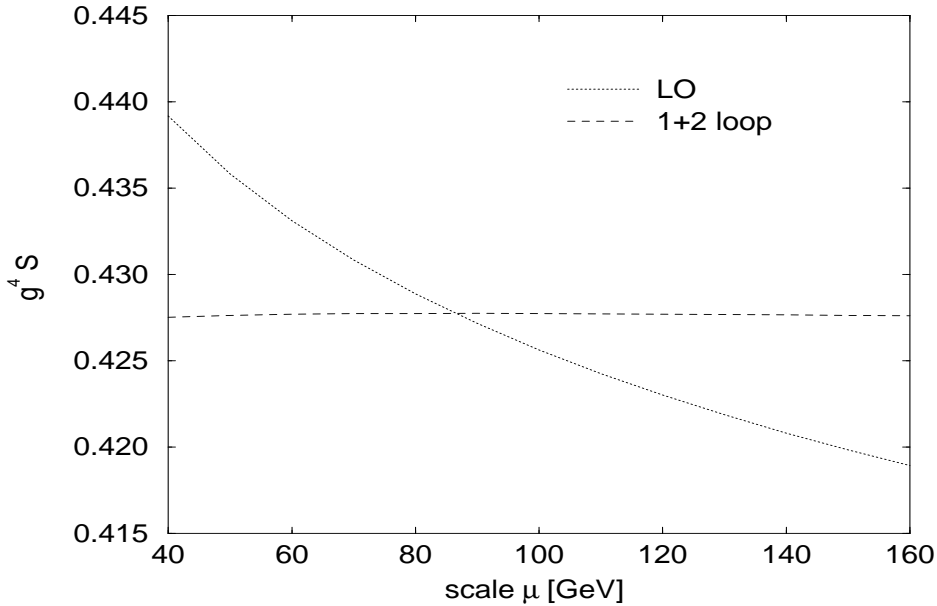}}}
\caption{\sf 
Dependence on the $\msbar$ scale of the LO and of the sum of LO and
two-loop results for $\mh=100$GeV. Inputs as in the text. }
\end{figure}
This dramatic reduction of the scale dependence
 demonstrates the importance of a complete two-loop \ew 
analysis in order to reduce the normalization ambiguities of the LO result.
\renewcommand{\arraystretch}{1.3}
\begin{table}[ht]
\[
\begin{array}{|c|  c| c |}\hline
{\rm scheme}
 & g^4 \, S_0 & g^4 \,(S_0+\delta S_{ew}) \\\hline \hline
\msbar & 0.4270 &  0.4277  \\ 
 G_\mu \mw^2  & 0.4320  & 0.4276   \\ 
{\rm OS}  &  0.4603 & 0.4270  \\ 
\hline
\end{array}            
\]
\caption{\sf Scheme dependence: comparison between LO and \ew corrected results
for $\mh=100$GeV.}
\end{table}

Following the discussion of Sec.\,3 on the normalization of the \ew coupling
$g$, we can also compare the  LO result $g^4 \,S_0(w_t)$
and the \ew corrected $g^4 \,(S_0(w_t)+\delta S_{ew})$ in different
renormalization schemes. Table 1 summarizes the scheme dependence before and
after the inclusion of our two-loop calculation by comparing 
the three representative cases introduced in Sec.\,3: 
$\msbar$ scheme at the scale $\mz$, 
$g$ expressed in terms of $G_\mu \mw^2$,
 $g$ expressed in terms of the {\it on-shell}
 sine $s_\smallw^2$ and of $\hat{\alpha}(\mz)$.  
The input values that we use are to a good approximation
compatible with the precise calculation summarized by the numerical 
formulas provided in Ref.\cite{DGPS} in the case $\mh=100$\,GeV and
 $\mt=174$\,GeV
(equivalent to $\mtbar=166$\,GeV). 
For our purposes it is therefore sufficient to evaluate numerically the extra 
terms $\delta S_{OS}$ and $\delta S_{G_\mu}$ inverting 
the definitions of $\Delta\hat{\rho}$ and $\Delta\hat{r}_\smallw$: one finds
$F_g^{OS}=(1-s_\smallw^2/\hat{s}^2(\mz))/2$ and 
$F_g^{G_\mu}= 1/2 \, (1-\pi\hat{\alpha}(\mz)/
(\sqrt{2} \,\hat{s}^2(\mz)\,G_\mu \,\mw^2))$, from which 
$\delta S_{OS}$ and $\delta S_{G_\mu}$ can be computed using
also $G_\mu=1.16639\times 10^{-5}$GeV$^{-2}$.
Here we have implicitly assumed that the  theoretical 
determination of the relations among the \ew couplings is ideally accurate;
although this is obviously not the case,  
the theoretical errors involved are of the order 
of a few parts in $10^4$ \cite{DGS} and can be neglected at this stage.

As could be expected  after 
the remarks made in Sec.~3 about the scheme dependence,
the difference between $\msbar$ scheme and $G_\mu$ normalization 
is only 1.1\% at LO, while the one between 
the $\msbar$ and {\it on-shell} schemes
is unnaturally very large for a purely \ew correction
at low-energies, almost 8\%. After the 
inclusion of our two-loop corrections, the situation changes drastically.
The maximum differences are now about $0.2\%$, which agrees with a 
rough estimate of the residual uncertainty, obtained  
by squaring the relative 
difference between $s^2_\smallw$ and $\hat{s}^2(\mz)$, that gives
$\approx 0.002$. 

Because  the {\it on-shell} scheme induces large and theoretically
well-controlled  radiative corrections $\delta S_{ew}$,  however,
it cannot be used to estimate higher order effects when the couplings are
normalized in a different way. We have  therefore extended 
the scheme comparison  of Table 1 to different values of $\mh$, 
limiting ourselves to the case where  the result is expressed in 
terms of $\msbar$ couplings or of $G_\mu$. 
To this end, one needs to recalculate $\hat{s}^2(\mz)$
for each $\mh$  from $G_\mu$ and $\mw$, and to  use the complete
expression for $F_g^{G_\mu}$. We find excellent agreement between the two
schemes: the residual ambiguity is always well below  $0.1\%$.

\subsection{Approximate formulae}
In this subsection we present some very simple approximate formulae that 
reproduce with excellent accuracy the result of our calculation when the LO
result is normalized to the $\msbar$ coupling $\hat{g}(\mu_g)$ and to 
$G_\mu \mw^2$. Unlike the original very long analytic formulae, they are
suitable for a simple implementation in numerical analyses. 

In the $\msbar$ scheme,
we can rewrite $\delta S_{ew}$  isolating the dependence of the complete 
result on the two scales $\mu_g$, at which the coupling $\hat{g} $ is
normalized, and $\mu_b\approx m_b$, at which the Wilson coefficient 
of the underlying effective theory is evaluated ($C(w_t)$ is the penguin
function given in \equ{Cfunc}):
\be
\delta S_{ew}^{\msbar}= \frac{\hat{g}^2}{16\pi^2} \left[
 S_0(w_t) \left(\frac{19}{3}
\,\ln \frac{\mu_g^2}{\mtbar^2} + \frac{2}{3} \hat{s}^2
 \,\ln \frac{\mu_b^2}{\mtbar^2}\right) + A(\mh,\mtbar) +16 \,C(w_t)^2\right].
\label{appr1}
\ee
The following simple approximate formula reproduces with high accuracy
the analytic  result  for $A(\mh,\mtbar)$, which cannot be reduced to  a 
 compact form:
\bea
A(\mh,\mtbar) &=&   12.20 + 52.74\,
    \left( {\frac{\mtbar}{166}}-1 \right)  - 
   0.977\,\ln {\frac{\mh}{100}} + 
   1.709\,\ln^2 \frac{\mh}{100},
\label{appr2}
\eea
where we have expressed the top and Higgs masses in GeV.
For $80<\mh<650$\,GeV, $\mtbar$ 
within $1\sigma$ from the present experimental value, $\mtbar=166\pm5$ GeV,
and  using $\hat{s}^2(\mz)=0.23145$ and $\mw=80.385$\,GeV, \equ{appr2}
does not deviate from the analytic result by more than 1\% which
induces in the complete result a very small $O(10^{-4})$ relative 
 error. 

\begin{figure}
\centerline{
\epsfysize=13truecm
\epsfxsize=8truecm
\rotate[r]{\epsffile{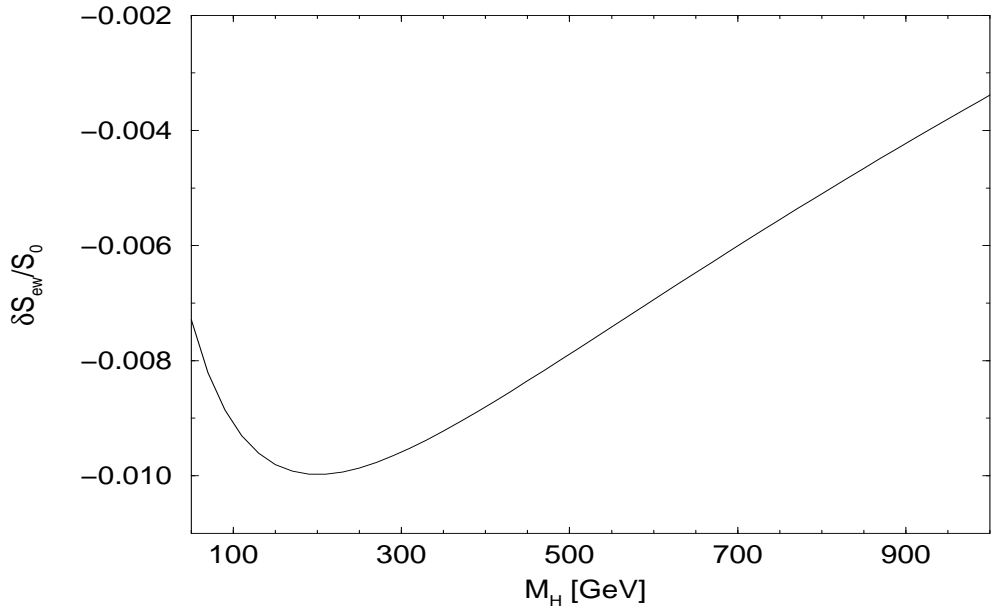}}}
\caption{\sf The \ew correction $\delta S_{ew}$ when the LO result is expressed
in terms of $G_\mu$, normalized to the LO result, as a function of the Higgs
mass. Inputs as in Fig.\,3.
}
\end{figure}
In the case the LO result is expressed in terms of $G_\mu\mw^2$, 
only the $\mu_b$ scale
dependence is left and  the Higgs and top dependence is  different due to 
the additional term $\delta S_{G_\mu}$, as shown in Fig.\,6.   The result can
then be written as 
\be
\delta S_{ew}^{G_\mu}= \frac{\sqrt{2}\,G_\mu\mw^2}{4\pi^2} \left[
 \frac{2}{3} \frac{\hat{\alpha}(\mz) \pi}{\sqrt{2}\,G_\mu\mw^2}
\,S_0(w_t) \,\ln \frac{\mu_b^2}{\mtbar^2} 
+ B(\mh,\mtbar) +16 \,C(w_t)^2\right],
\label{appr3}
\ee
where
\bea
B(\mh,\mtbar)
&=&  -15.49 -31.65 \left(\frac{\mtbar}{166}-1\right) - 
   2.296 \,\ln\frac{\mh}{100} + 1.868\,\ln^2\frac{\mh}{100}
\label{appr4},
\eea
and again $\mtbar$ and $\mh$ are expressed in GeV. In the same ranges of 
$\mtbar$ and $\mh$ that we
considered for the function $A(\mh,\mtbar)$, $B(\mh,\mtbar)$ approximates
the complete result with analogous accuracy.

\section{Conclusions}
In this paper we have computed the complete \ew effects in the $\bb$ mixing
in the SM.
We have used an effective theory approach in treating the QED corrections
and neglected QED effects in  the matrix elements. 
The calculation has been performed analytically, and the results expressed 
for arbitrary Higgs mass in terms of the very simple 
approximate formulae of Eqs.\,(\ref{appr1}--\ref{appr4}).
The discussion of Sec.\,4 can be summarized in the following way:
\begin{enumerate}
\item Unless the renormalization procedure introduces 
unnaturally large correction unrelated to the process, the two-loop 
\ew corrections are  small, $O(1\%)$, for any realistic 
value of the Higgs mass.  This is  of the same 
order of magnitude of the NLO QCD corrections. 
If one uses an {\it on-shell}
definition of the \ew coupling, however, they reach almost 8\% for a light
Higgs boson.

\item The \ew 
scheme and scale dependence of the result, which we have shown to be
quite large at LO, is  consistently reduced to the permille level.
If $G_\mu \mw^2$ or $\msbar$ couplings are used
to normalize the result, the ambiguity 
is  below 0.1\%, much less than the present perturbative  QCD error.

\item The heavy top expansion converges very slowly in the case of the $\bb$
mixing. This was known at LO and we have verified the same pattern
 at the two-loop level.
The fact that the {\it leading} heavy top limit
 does {\it not}  approximate well  the complete result 
suggests  that it cannot be used as a reliable estimate of the full \ew
effects in cases where already at LO it does not work properly.
In particular, this refers to some weak decays for which the leading heavy top
corrections have recently been computed:   
see \cite{strumia:98}
(for the decay $B \rightarrow X_s \gamma$)  and \cite{buchalla:98}
(for a class of rare  decays including $K\to\pi \nu\bar{\nu}$).
\end{enumerate}

If one expresses the LO result in terms of  $G_\mu$ and $\mw$, as it is 
customarily done
\cite{buchalla:96}, the correction as a function of the Higgs mass
is shown in Fig.\,6 and 
 the effect on the extraction of $|V_{td}|$ from $\Delta M_{B_d}$
is tiny. Indeed, $|V_{td}|\sim (S_0+\delta S_{ew})^{-1/2}$, so 
 the \ew corrections induce approximately a $+0.5\%$ change in the
extracted value of $|V_{td}|$, for values of the Higgs mass below 400\,GeV.

\subsection*{Acknowledgments}
We are grateful to A.J. Buras for suggesting this project to us and  for
interesting discussions. We also thank M. Misiak for carefully reading  the
manuscript and  
P.A. Grassi,  T. Hahn,  L. Silvestrini, and M. Steinhauser
 for useful communications and discussions. One of us (N.P.) would
like to thank W. Kistler for his help with several computer problems.

\end{document}